\documentclass[aps,prl,twocolumn,preprintnumbers,superscriptaddress,amsmath,floatfix]{revtex4}
\usepackage{amssymb}
\usepackage{txfonts}
\usepackage{amssymb}
\usepackage{graphicx}
\usepackage{subfigure}
\usepackage{dcolumn}
\usepackage{bm}
\usepackage{color}
\usepackage{upgreek}

\begin{document}


\title{Single-photon non-reciprocity with an integrated magneto-optical isolator}
\author{Shang-Yu Ren}
\thanks{These authors contributed equally to this work.}
\affiliation{Key Laboratory of Quantum Information, University of Science and Technology of China, Hefei 230026, China
}
\affiliation{
Synergetic Innovation Center of Quantum Information and Quantum Physics, University of Science and Technology of China, CAS, Hefei 230026, China	
}
\author{Wei Yan}
\thanks{These authors contributed equally to this work.}
\affiliation{National Engineering Research Center of Electromagnetic Radiation Control Materials, University of Electronic Science and Technology of China, Chengdu 610054, China
}
\affiliation{State Key Laboratory of Electronic Thin-Films and Integrated Devices, University of Electronic Science and Technology of China, Chengdu, 610054, China
}
\author{Lan-Tian Feng}
\thanks{These authors contributed equally to this work.}
\affiliation{Key Laboratory of Quantum Information, University of Science and Technology of China, Hefei 230026, China
}
\affiliation{
Synergetic Innovation Center of Quantum Information and Quantum Physics, University of Science and Technology of China, CAS, Hefei 230026, China	
}
\author{Yang Chen}
\author{Yun-Kun Wu}
\author{Xiao-Zhuo Qi}
\author{Xiao-Jing Liu}
\author{Yu-Jie Cheng}
\author{Bo-Yu Xu}
\affiliation{Key Laboratory of Quantum Information, University of Science and Technology of China, Hefei 230026, China
}
\affiliation{
Synergetic Innovation Center of Quantum Information and Quantum Physics, University of Science and Technology of China, CAS, Hefei 230026, China	
}
\author{Long-Jiang Deng}
\affiliation{National Engineering Research Center of Electromagnetic Radiation Control Materials, University of Electronic Science and Technology of China, Chengdu 610054, China
}
\affiliation{State Key Laboratory of Electronic Thin-Films and Integrated Devices, University of Electronic Science and Technology of China, Chengdu, 610054, China
}
\author{Guang-Can Guo}
\affiliation{Key Laboratory of Quantum Information, University of Science and Technology of China, Hefei 230026, China
}
\affiliation{
Synergetic Innovation Center of Quantum Information and Quantum Physics, University of Science and Technology of China, CAS, Hefei 230026, China		
}
\author{Lei Bi}
\email{bilei@uestc.edu.cn}
\affiliation{National Engineering Research Center of Electromagnetic Radiation Control Materials, University of Electronic Science and Technology of China, Chengdu 610054, China
}
\affiliation{State Key Laboratory of Electronic Thin-Films and Integrated Devices, University of Electronic Science and Technology of China, Chengdu, 610054, China
}
\author{Xi-Feng Ren}
\email{renxf@ustc.edu.cn}
\affiliation{Key Laboratory of Quantum Information, University of Science and Technology of China, Hefei 230026, China
}
\affiliation{
Synergetic Innovation Center of Quantum Information and Quantum Physics, University of Science and Technology of China, CAS, Hefei 230026, China	
}
\date{\today}
\begin{abstract}
Non-reciprocal photonic devices are essential components of classical optical information processing. It is interesting and important to investigate their feasibility in the quantum world. In this work, a single-photon non-reciprocal dynamical transmission experiment was performed with an on-chip silicon nitride (SiN)-based magneto-optical (MO) isolator. The measured isolation ratio for single photons achieved was 12.33 dB, consistent with the result of classical test, which proved the functionality of our on-chip isolator. The quantum coherence of the passing single photons was further verified using high-visibility quantum interference. Our work will promote on-chip nonreciprocal photonic devices within the integrated quantum circuits and help introduce novel phenomena in quantum information processes.
\end{abstract}

\maketitle
\section{Introduction}
Photonic integrated circuits (PICs) possess inherent advantages over space and fiber optics owing to their small footprint, scalability, programmability, and stability against the environment \cite{RN20,RN59,RN67,ZhangJie-150}. These advantages also become quite important for large scale quantum information process \cite{ReckZeilinger-139,Politi,Feng}. As quantum information processing becomes much more complicated, requiring larger number of optical components \cite{Pan}, the demand for quantum PICs (QPICs) will significantly increase. Although many photonic elements, such as phase shifters, directional couplers, and even high-Q cavities, have already been integrated on QPICs, however, till date, the non-reciprocal optical isolator has not been used for on-chip quantum circuits. 

Non-reciprocal photonic devices, which break the time-reversal symmetry, play key roles in optical information processing. They work as essential isolators and circulators to protect the laser source and eliminate the Fabry–Perot effect. In quantum information processing, non-reciprocal devices can be used to explore quantum transfer in optical systems \cite{Mascarenhas}. Further, by designing networks for non-reciprocal devices, it is possible to construct directional, quantum-limited amplifiers \cite{Metelmann}.

Many non-reciprocal devices have been reported in bulk structures \cite{zhang,Hu}, and nanostructures \cite{Shen,Chai} and are widely used in commercial optical fibers. Recently, integrated optical isolators based on magneto-optical effects \cite{Bi, Yan, Zhang:19, Pintus:19}, nonlinear photonic effects \cite{Fan447, Wang:20, RenXu-39} and spatio-temporal modulation \cite{KittlausOtterstrom-40, Sohn, Liang} have also been reported. Among them, Silicon-On-Insulator (SOI) and SiN-based MO isolators are particularly promising owing to their wide optical isolation bandwidth, low insertion loss, and passive operation capability \cite{Yan,Zhang:19}. This provides us an opportunity to integrate the isolator onto a quantum chip. Moreover, the permittivity tensor of the MO materials allows a linear relationship between the polarization P($\omega$) and the incident electric field E($\omega$). However, whether this property remains valid or not to the single-photon condition is still unknowns. 

Here, by taking a single-photon-level transmission test, we prove the functionality of an on-chip magneto-optical isolator with an isolation ratio of 12.33 dB at 1550 nm wavelength. Further, the Hong-Ou-Mandel (HOM) interference experiments with high visibility also indicate that the coherence of transmitted photons can be well preserved. These results reveal the potential utility of non-reciprocal photonic devices in QPICs.

\section{Experimental Section}
\subsection{Fabrication and classical characterization of the magneto-optical isolator}
\begin{figure}[htbp]
	\vspace{-0cm}
	\centering
	\includegraphics[scale=0.37]{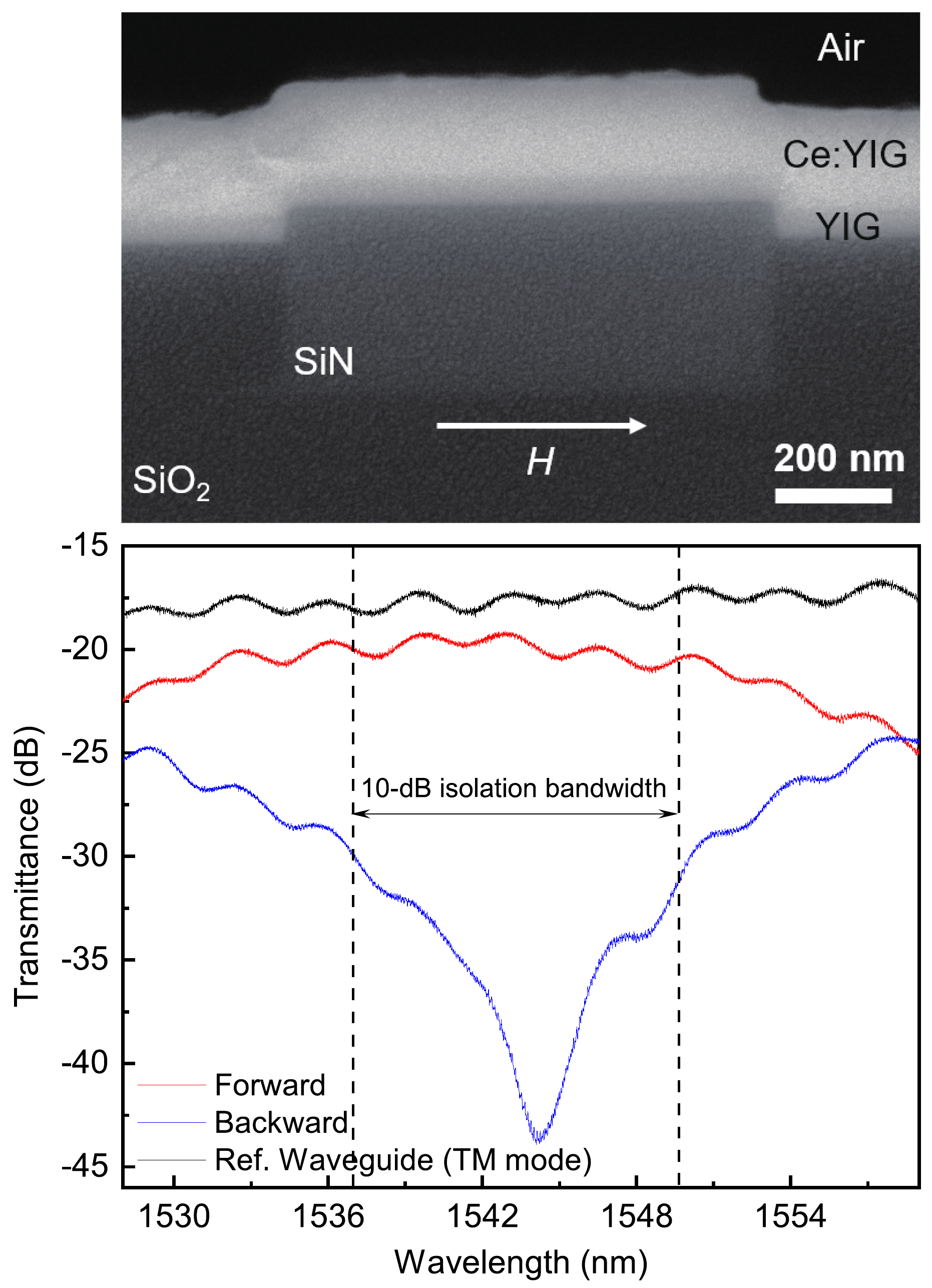}
	\caption{(a) Scanning electron microscopy (SEM) image of the cross section of a fabricated MO/SiN waveguide in the SiN optical isolator. (b) Classically measured transmission spectra of the TM mode SiN optical isolators with a TM mode SiN reference waveguide.}
	\vspace{-0cm}
	\label{经典测试图}
\end{figure}

We began by fabricating a SiN transverse magnetic (TM) MO isolator based on the Mach-Zehnder interferometer (MZI) structure \cite{Zhang:19, Shoji, Huang}, using standard silicon photonics foundry processes for fabricating SiN PICs and pulsed laser deposition (PLD) for MO thin films. 3 dB directional couplers were introduced at both ends of the MZI. The reciprocal and non-reciprocal phase shifters on both arms of the MZI allowed a 0 ($\pi$) phase difference for the forward (backward) propagating light, thus achieving isolation. The non-reciprocal phase shifter was an MO/SiN waveguide with Ce:YIG/YIG thin films on top of the SiN waveguide, providing a non-reciprocal phase shift (NRPS) for TM-polarized light \cite{Pintus:19,Huang:17}. Fig.~\ref{经典测试图}(a) shows the cross-sectional scanning electron microscope (SEM) image of the fabricated MO/SiN waveguide in the SiN device. The structure consisted of a planar Ce:YIG (150 nm)/YIG (50 nm) thin-film stack deposited on the SiN channel waveguide. The MO/SiN waveguide in each arm of the MZI structure was designed to be 670 $\mu$m in length, as mentioned in \cite{Yan}. The Faraday rotation of YIG and Ce:YIG thin films was set as 500 deg/cm and -5900 deg/cm, respectively. The refractive index of the MO thin film was 2.3. During the classical measurement, an in-plane magnetic field (1000 Gs) was applied perpendicular to the propagation direction of light to saturate the magnetization of the MO thin films.
Fig.~\ref{经典测试图}(b) shows the transmission spectra of the TM mode SiN optical isolator, together with a reference SiN waveguide (TM mode) on the same chip. At 1544 nm wavelength, the maximum isolation ratio reached 25 dB, with an insertion loss of 2.3 dB. The 10 dB isolation bandwidth of this device was 13 nm, and across the entire 10 dB isolation bandwidth, the device experienced an insertion loss of 2.3-3 dB. The isolation ration decreased to 11 dB at 1550 nm wavelength with an insertion loss increasing to 3 dB.
\begin{figure*}[bht]
	\vspace{-0cm}
	\centering
	\includegraphics[scale=0.57]{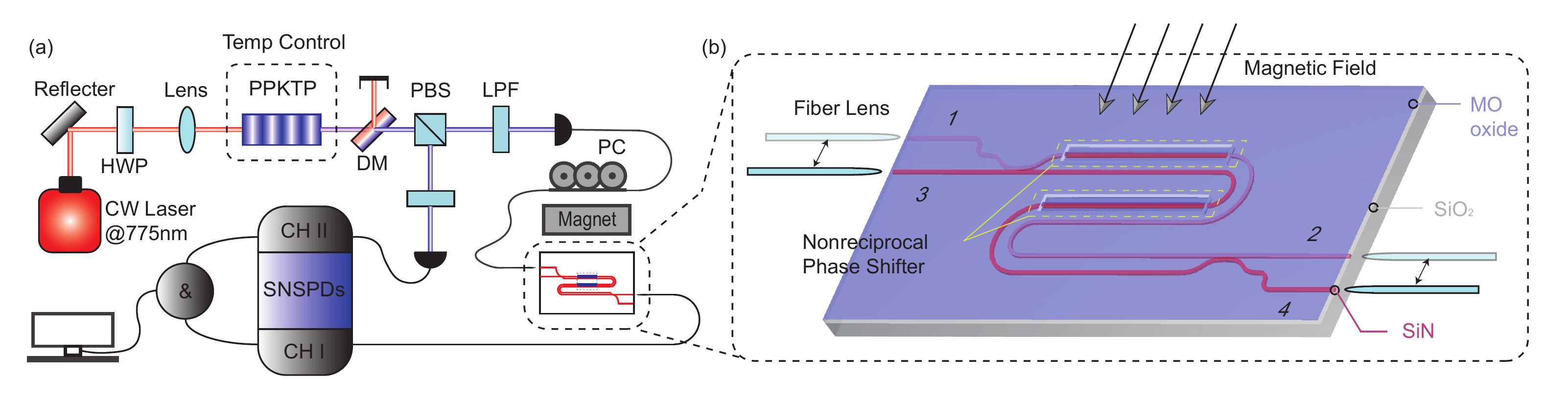}
	\caption{Schematic of the experimental setup and non-reciprocal optical resonator structure. (a) The experimental setup. HWP: half-wave plate, used to adjust the input polarization; DM: dichroic mirror to filter out the pump laser; PPKTP: nonlinear crystal, used to generate frequency-degraded photon pairs; PBS: polarization beam splitter, used to split photons pairs into different paths. LPF: long-pass filter to filter out the residual pump laser; SNSPDs: superconducting nanowire single-photon detectors for coincidence measurement. PC: the polarization controller is also used to maximize the coupling efficiency. (b) Schematic of the non-reciprocal optical resonator structure.}
	\vspace{-0cm}
	\label{实验装置图}
\end{figure*} 
\subsection{Experiment of single-photon non-reciprocal transmission}
A schematic of the experimental setup for the quantum test is shown in Fig.~\ref{实验装置图}. A continuous-wave laser centered at 775 nm was used as the pump (Toptical DL Pro) to generate photon pairs at 1550 nm through a spontaneous parametric down-conversion (SPDC) process in a 3 cm long potassium titanyl phosphate (PPKTP) crystal. The temperature of the crystal was controlled to achieve frequency-degenerate photon pairs and the wavelength was fixed via a wavelength meter (Angstrom WS/6 IR; HighFinesse). The generated photon pairs had orthogonal polarizations; thus, they were divided into two paths by a polarization beam splitter (PBS) and then coupled into SM-28 single-mode fibers. Fiber lenses (Chuxing Optical Fiber Application Technologies Ltd) were used for the on-chip coupling. In our experiment, one photon of each pair was coupled into and passed through the on-chip isolator. The other photon was directly inserted into one detector and acted as a reference. Both photons were detected by superconducting nanowire single-photon detectors (SNSPDs, Scontel). The output electric signals were imported into the TCSPC system for the correlation analysis.
\begin{figure}[tbp]
	\vspace{-0cm}
	\centering
	\includegraphics[scale=0.4]{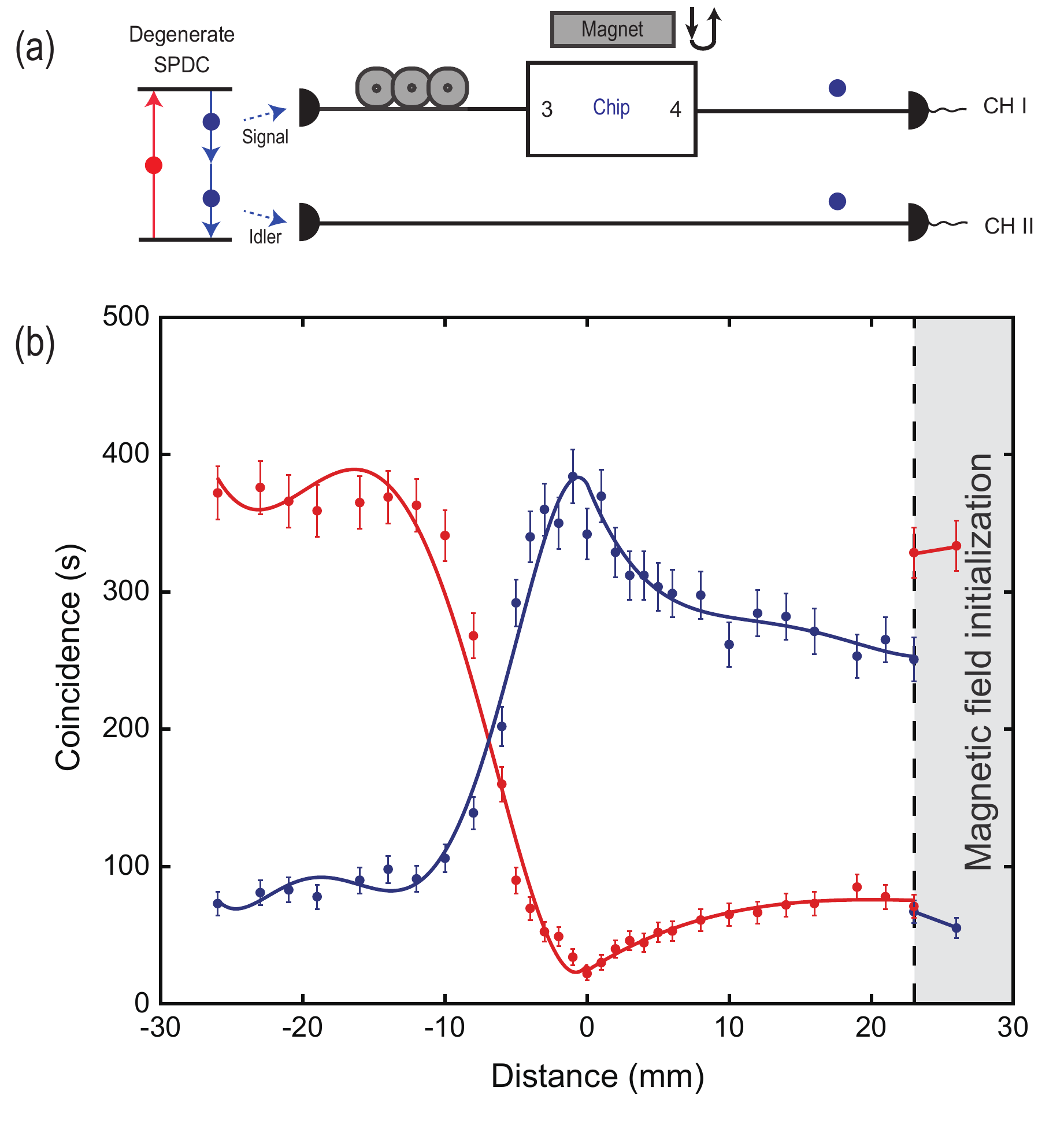}
	\caption{(a) Experimental setup of single-photon nonreciprocal transmission test. The magnet approaches and then moves away from the chip to observe the variation of coincidence counts with the strength of magnetic field. (b)The measured coincidence. Blue dots: for Case $ B $, signal photons pass from port 3 to port 4. Red dots: for Case $ B^\prime $, signal photons pass from port 4 to port 3. Negative abscissa axis indicates the process of magnet approaching the chip. Positive abscissa axis indicates the process of the magnet moving away from the chip. The curves only show the changing trend of the coincidence counts. Error bar comes from the Poisson statistical distribution. The initial magnetization is opposite to the magnetic field direction of the moving magnet. Finally, the magnetic field in chip was initialized again in the grey zone.}
	\vspace{-0cm}
	\label{单光子符合图}
\end{figure}
The core of the on-chip MO isolator is the MZI with four ports. We studied four propagation situations of the signal photons. $ A $: From port 1 to 2, $ A^\prime $: from port 2 to 1, $ B $: from port 3 to 4, and $ B^\prime $: from port 4 to 3 of the isolator. The polarization of the signal photons must be fixed by a polarization controller to excite them into a TM mode to match the working polarization of the device because only the TM mode induces NRPS under the classical test. A magnet was placed parallel to the device, producing an in-plane external magnetic field perpendicular to the propagation direction of the light. The magnitude of the generated magnetic field affects the strength of the non-reciprocal phase shift effect.

In the experiment (Fig.~\ref{单光子符合图}(a)), the magnet was fixed on a self-made scaled displacement stage, which could move forward and backward to observe the influence of the magnetic field magnitude on single-photon transmission. The signal photons passed through the chip and were received by Channel I of the SNSPDs. The idler photons were directly received by Channel II, and the coincidences between the two channels were recorded. Fig.~\ref{单光子符合图}(b) shows the experimental results. The horizontal axis coordinates denote the distance between the magnet and the end face of the chip. The blue dots indicate the measured coincidence counts for Case B. The red dots denote the case in which we reversed the direction of propagation of the photons, that is, Case $ B^\prime $. 

At the beginning of the test, the magnet was placed 26 mm away from the chip. This distance guaranteed that the magnetic field had no influence on the MO materials. Focusing on the red curve (Case $B^\prime $), when the magnet is moving from -26 mm to -10 mm, the transmission is steady because the external magnetic field is still too small to induce an obvious non-reciprocal phase shift effect. When the distance between the magnet and the device reduces sufficiently, the number of transmitted photons decreases quickly. At the saturation magnetization point (0 distance), the coincidence reaches a minimum. When the magnet is pulled back, the coincidence increases but is not as fast because of the magnetic remanence magnetization in the MO materials. For the blue curve (Case $B$), the phenomenon is the opposite. These two curves present non-reciprocal characteristics. It is worth noting that we applied a magnetic field initialization operation when the magnet was pulled back to the original location after the saturation magnetization because of remanent magnetization in the MO thin film. The coincidence rises or falls sharply and returns to the initial value after this operation (shown in the gray zone in Fig.~\ref{单光子符合图}(b)).

Tests for Cases $A$ and $A^\prime $ were also conducted and the results were consistent with those in the Case $B$ and $B^\prime $. For comparison, we also measured the coincidence for photons passing through a straight SiN waveguide, in contrast to the isolator (see Supplementary Material). The result proves that the trend of coincidence is not caused by the magnet's effect on the coupling. The maximum isolation ratio for the single photons was approximately 12.33 dB for $B^\prime $. This is lower than that in classical characterization at 1544 nm wavelength, because our source needed to be fixed at 1550 nm wavelength. Even though it is not the best working wavelength of the isolator, the result was basically consistent with the classical one in Section 2.1 at 1550 nm wavelength.

\subsection{HOM experiment on the magneto-optical non-reciprocal isolator}
\begin{figure}[t]
	\vspace{-0cm}
	\centering
	\includegraphics[scale=0.4]{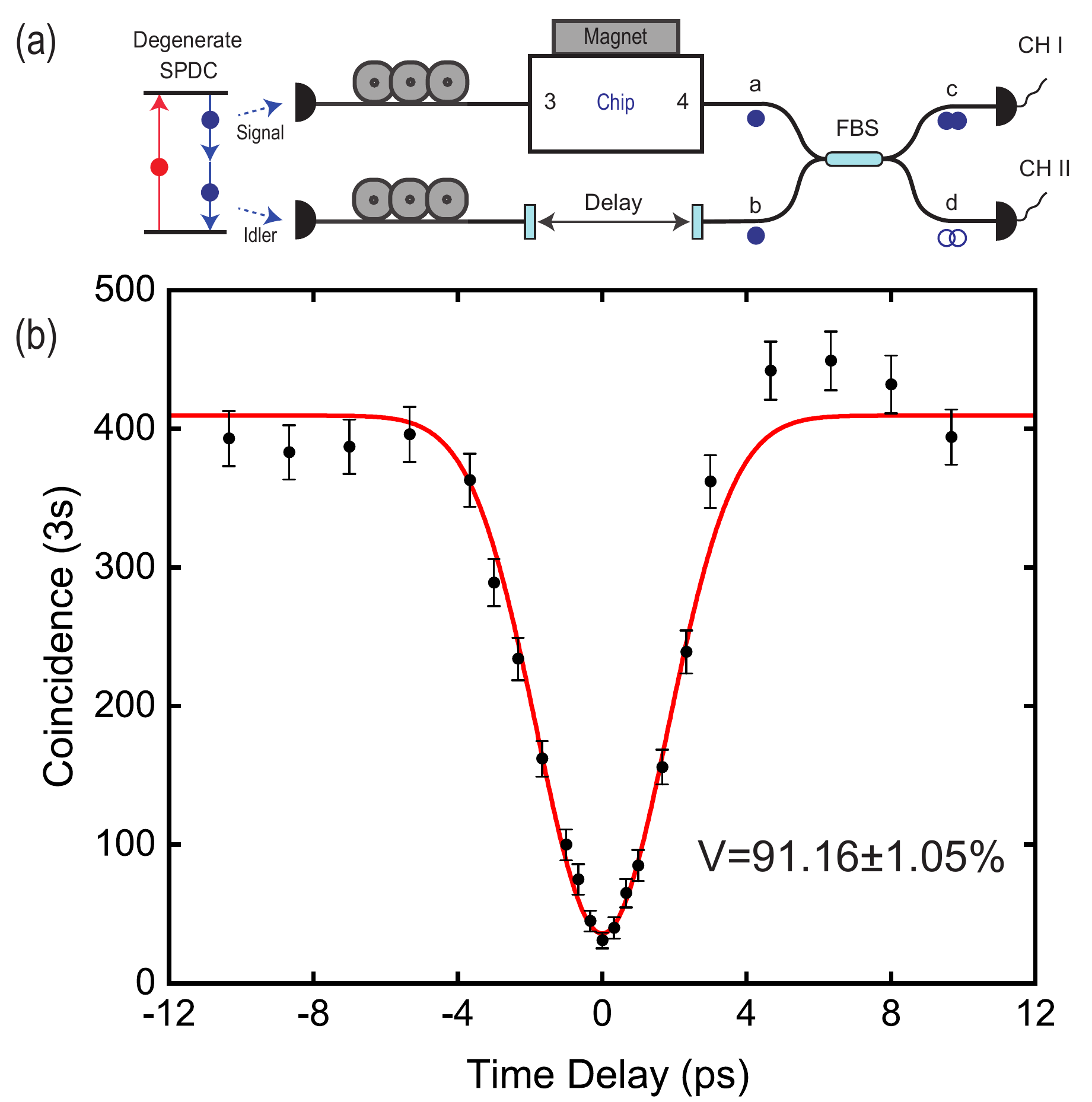}
	\caption{(a) Experimental setup of the off-chip HOM interference. Magnet is placed next to the chip to obtain saturation magnetization of the MO thin film. (b) Coincidence when the signal photons pass from port 1 to port 2 of the chip. Accidental coincidences have been subtracted. The red curve is fitted using the Lorentz function. Error bar comes from the Poisson statistical distribution based on raw coincidences.}
	\vspace{-0cm}
	\label{HOM干涉图}
\end{figure}

To prove the validity of the device for QPICs, we also need to verify whether the coherence of the transmitted photons can be preserved. To achieve this, the HOM interference was used, which is a basic interference in quantum physics that can reflect the bosonic properties of a single particle. It is typically used to test the quantum properties of single qubits \cite{HOM}. Experimentally, a 50/50 fiber beam splitter (FBS) was used to perform the interference (Fig.~\ref{HOM干涉图}(a)). A single photon passed through the chip and the reference photon is imported into the FBS at ports a and b. We can derive the interference process using the transformation matrix of the FBS:
\begin{equation}
	\begin{aligned}
		a^\dagger b^\dagger\left|0\right \rangle &\overset{FBS}{\longrightarrow}\frac{1}{2}(c^\dagger+d^\dagger)(c^\dagger-d^\dagger)\left|0\right \rangle\\
		&=\frac{1}{2}(c^{\dagger2}-c^\dagger d^\dagger+d^\dagger c^\dagger-d^{\dagger2})\left|0\right \rangle\\
		&=\frac{1}{2}(c^{\dagger2}-d^{\dagger2})\left|0\right \rangle\\
	\end{aligned}
\end{equation}
$a^\dagger$, $b^\dagger$, $c^\dagger$ and $d^\dagger$ are creation operators at corresponding ports of FBS. Only when photons are indistinguishable, $-c^\dagger d^\dagger+d^\dagger c^\dagger$ can be eliminated. The last line of the formula means the photons can be detected either at c or d port of FBS with one-half probability respectively.

Therefore, in this case, there is no coincidence between ports c and d of the FBS. Without loss of generality, we assume that the spectrum transmitted by the interference filter is Gaussian. The coincidence count can be expressed as $ C\propto1-e^{-\Delta\tau^2\Delta w^2} $, where $\Delta\tau$ refers to the time delay of two photons on the FBS and $ \Delta w $ is the spectral width of the photons. In the experiment, the polarization controllers were used to ensure that the photons had the same polarization before interfering in the FBS. We synchronized the arrival time of the photon pair by adjusting the path-length difference between them. The time delay was introduced by a group of two fiberport collimators.

The visibility (\textit{V}) of the HOM interference is defined as $V=(C_{max}-C_{min})/C_{max}$, where $C_{max}$ is the maximum coincidence and $C_{min}$ is the minimum coincidence. When $C_{min}=0$, the perfect coincidence visibility is $V=1$, which indicates that the two photons are identical. Before testing on the non-reciprocal isolator, we characterized our SPDC source. The HOM interference visibility is up to 95.61±0.94$\%$. The experimentally measured interference visibility in Case $B$ was 91.16±1.05$\%$ , which indicates that the photons maintain their coherence after the function of the MO isolator. HOM interference tests for the source, waveguide and other cases have been discussed in the supplementary material.

\section{Discussion}
In this work, we didn't use any additional filters to manipulate the bandwidth of SPDC photon pair because PPKTP crystal itself played a filtering role. However, the bandwidth of quasi phase matching  in PPKTP crystal is still over 100 GHz. That means althought the single-photon isolation is consistent with the result of classical test, it is actually determined by the combined effect of the photons around 1550 nm. As described above, 3 dB directional couplers were introduced to the characterized device at both ends of the MZI structure, forming a four-port optical circulator. Therefore, the device we showed above can also achieve the function of optical circulation. The classical characterization of the optical circulator can be found in \cite{Yan}. Although the device worked under the TM polarization which mismatched most integrated photonic devices designed for the TE polarization, TE-TM polarization rotators (PRs) can be introduced at both ends of the TM devices for constructing optical isolators and circulators operating for the TE polarization \cite{Pintus:19}. To minimize the insertion loss, polarization rotators based on the mode evolution mechanism to rotate the fundamental TE mode into the fundamental TM mode have been practical realized \cite{Yin:17}. There is also an issue that cannot be ignored on the edge coupling loss during the test, both classsically and quantumly. The high fiber-to-chip test loss was mainly caused by the high edge coupling loss, which was over 5 dB for each end. It can be further optimized by designing trident spot-size converter (SSC) \cite{Hatori} combining three misplaced tapers or other high-efficiency couplers with air trenches \cite{8626148}. 

\section{Conclusion}
Our experiment proved unambiguously that the fabricated on-chip isolator works well at the single-photon level. In addition, the quantum properties of photons passing through the device are well maintained. Our work shows the potential of the integrated non-reciprocal device as an on-chip isolator and for its use in quantum algorithms and quantum computing circuits. For example, it can be integrated with on-chip lasers \cite{Xiang:20} and other photonic components such as filters \cite{Harris,LiuDajian}, photon sources \cite{Jin2} and quantum circuits. It can serve as a feasible photonic isolator, and join and support the full on-chip generation of arbitrary quantum states \cite{Paesani}, teleportation \cite{Metcalf} and quantum computation \cite{Knill}.

\section{Acknowledgements}
S.-Y.R., W.Y. and L.-T.F. contributed equally to this work. The work is supported by National Natural Science Foundation of China (NSFC) (Grants No. 61590932, No. 11774333, No. 62061160487, No. 12004373, No. 51972044), Ministry of Science and Technology of the People's Republic of China (MOST) (Grant No. 2018YFE0109200), the Anhui Initiative in Quantum Information Technologies (Grant No. AHY130300), the Strategic Priority Research Program of the Chinese Academy of Sciences (Grant No. XDB24030601), the National Key R\&D Program (Grant No. 2016YFA0301700), the Postdoctoral Science Foundation of China (Grant No. 2020M671860) and the Fundamental Research Funds for the Central Universities. This work was partially carried out at the USTC Center for Micro and Nanoscale Research and Fabrication.
\bibliography{sn_MK5}

\begin{thebibliography}{10}
\providecommand{\url}[1]{\texttt{#1}}
\providecommand{\urlprefix}{URL }

\bibitem{RN20}
M.~Zhang, C.~Wang, Y.~Hu, A.~Shams-Ansari, T.~Ren, S.~Fan, M.~Lončar,
\newblock \emph{Nat. Photonics} \textbf{2018}, \emph{13}, $\!\!$ 36.

\bibitem{RN59}
M.~Li, H.~Liang, R.~Luo, Y.~He, J.~Ling, Q.~Lin,
\newblock \emph{Optica} \textbf{2019}, \emph{6}, $\!\!$ 860.

\bibitem{RN67}
B.~Machielse, S.~Bogdanovic, S.~Meesala, S.~Gauthier, M.~J. Burek, G.~Joe,
  M.~Chalupnik, Y.~I. Sohn, J.~Holzgrafe, R.~E. Evans, C.~Chia, H.~Atikian,
  M.~K. Bhaskar, D.~D. Sukachev, L.~Shao, S.~Maity, M.~D. Lukin,
  M.~Lon\ifmmode~\check{c}\else \v{c}\fi{}ar,
\newblock \emph{Phys. Rev. X} \textbf{2019}, \emph{9}, $\!\!$ 031022.

\bibitem{ZhangJie-150}
L.~Zhang, L.~Jie, M.~Zhang, Y.~Wang, Y.~Xie, Y.~Shi, D.~Dai,
\newblock \emph{Photonics Research} \textbf{2020}, \emph{8}, $\!\!$ 684.

\bibitem{ReckZeilinger-139}
M.~Reck, A.~Zeilinger, H.~J. Bernstein, P.~Bertani,
\newblock \emph{Phys. Rev. Lett.} \textbf{1994}, \emph{73}, $\!\!$ 58.

\bibitem{Politi}
A.~Politi, J.~C.~F. Matthews, J.~L. O'Brien,
\newblock \emph{Science} \textbf{2009}, \emph{325}, $\!\!$ 1221.

\bibitem{Feng}
L.-T. Feng, M.~Zhang, Z.-Y. Zhou, M.~Li, X.~Xiong, L.~Yu, B.-S. Shi, G.-P. Guo,
  D.-X. Dai, X.-F. Ren, G.-C. Guo,
\newblock \emph{Nat. Commun.} \textbf{2016}, \emph{7}, $\!\!$ 11985.

\bibitem{Pan}
H.~S. Zhong, H.~Wang, Y.~H. Deng, M.~C. Chen, L.~C. Peng, Y.~H. Luo, J.~Qin,
  D.~Wu, X.~Ding, Y.~Hu, P.~Hu, X.~Y. Yang, W.~J. Zhang, H.~Li, Y.~Li,
  X.~Jiang, L.~Gan, G.~Yang, L.~You, Z.~Wang, L.~Li, N.~L. Liu, C.~Y. Lu, J.~W.
  Pan,
\newblock \emph{Science} \textbf{2020}, \emph{370}, $\!\!$ 1460.

\bibitem{Mascarenhas}
E.~Mascarenhas, F.~Damanet, S.~Flannigan, L.~Tagliacozzo, A.~J. Daley,
  J.~Goold, I.~De~Vega,
\newblock \emph{Phy. Rev. B} \textbf{2019}, \emph{99}, $\!\!$ 245134.

\bibitem{Metelmann}
A.~Metelmann, A.~A. Clerk,
\newblock \emph{Phys. Rev. X} \textbf{2015}, \emph{5}, $\!\!$ 021025.

\bibitem{zhang}
R.-Y. Zhang, Y.-W. Zhai, S.-R. Lin, Q.~Zhao, W.~Wen, M.-L. Ge,
\newblock \emph{Sci. Rep.} \textbf{2015}, \emph{5}, $\!\!$ 1.

\bibitem{Hu}
X.-X. Hu, Z.-B. Wang, P.~Zhang, G.-J. Chen, Y.-L. Zhang, G.~Li, X.-B. Zou,
  T.~Zhang, H.~X. Tang, C.-H. Dong, et~al.,
\newblock \emph{Nat. Commun.} \textbf{2021}, \emph{12}, $\!\!$ 1.

\bibitem{Shen}
Z.~Shen, Y.-L. Zhang, Y.~Chen, C.-L. Zou, Y.-F. Xiao, X.-B. Zou, F.-W. Sun,
  G.-C. Guo, C.-H. Dong,
\newblock \emph{Nat. Photonics} \textbf{2016}, \emph{10}, $\!\!$ 657.

\bibitem{Chai}
C.~Z. Chai, H.~Q. Zhao, H.~X. Tang, G.~C. Guo, C.~L. Zou, C.~H. Dong,
\newblock \emph{Laser Photonics Rev.} \textbf{2020}, \emph{14}, $\!\!$ 1900252.

\bibitem{Bi}
L.~Bi, J.~Hu, P.~Jiang, D.~H. Kim, G.~F. Dionne, L.~C. Kimerling, C.~A. Ross,
\newblock \emph{Nat. Photonics} \textbf{2011}, \emph{5}, $\!\!$ 758.

\bibitem{Yan}
W.~Yan, Y.~Yang, S.~Liu, Y.~Zhang, S.~Xia, T.~Kang, W.~Yang, J.~Qin, L.~Deng,
  L.~Bi,
\newblock \emph{Optica} \textbf{2020}, \emph{7}, $\!\!$ 1555.

\bibitem{Zhang:19}
Y.~Zhang, Q.~Du, C.~Wang, T.~Fakhrul, S.~Liu, L.~Deng, D.~Huang, P.~Pintus,
  J.~Bowers, C.~A. Ross, J.~Hu, L.~Bi,
\newblock \emph{Optica} \textbf{2019}, \emph{6}, $\!\!$ 473.

\bibitem{Pintus:19}
P.~Pintus, D.~Huang, P.~A. Morton, Y.~Shoji, T.~Mizumoto, J.~E. Bowers,
\newblock \emph{J. Lightwave Technol.} \textbf{2019}, \emph{37}, $\!\!$ 1463.

\bibitem{Fan447}
L.~Fan, J.~Wang, L.~T. Varghese, H.~Shen, B.~Niu, Y.~Xuan, A.~M. Weiner, M.~Qi,
\newblock \emph{Science} \textbf{2012}, \emph{335}, $\!\!$ 447.

\bibitem{Wang:20}
J.~Wang, Y.~Shi, S.~Fan,
\newblock \emph{Opt. Express} \textbf{2020}, \emph{28}, $\!\!$ 11974.

\bibitem{RenXu-39}
L.~Ren, X.~Xu, S.~Zhu, L.~Shi, X.~Zhang,
\newblock \emph{ACS Photonics} \textbf{2020}, \emph{7}, $\!\!$ 2995.

\bibitem{KittlausOtterstrom-40}
E.~A. Kittlaus, N.~T. Otterstrom, P.~Kharel, S.~Gertler, P.~T. Rakich,
\newblock \emph{Nat. Photonics} \textbf{2018}, \emph{12}, $\!\!$ 613.

\bibitem{Sohn}
D.~B. Sohn, S.~Kim, G.~Bahl,
\newblock \emph{Nat. Photonics} \textbf{2018}, \emph{12}, $\!\!$ 91.

\bibitem{Liang}
C.~Liang, B.~Liu, A.-N. Xu, X.~Wen, C.~Lu, K.~Xia, M.~K. Tey, Y.-C. Liu,
  L.~You,
\newblock \emph{Phys. Rev. Lett.} \textbf{2020}, \emph{125}, $\!\!$ 123901.

\bibitem{Shoji}
Y.~{Shoji}, A.~{Fujie}, T.~{Mizumoto},
\newblock \emph{IEEE J. Sel. Top. Quant.} \textbf{2016}, \emph{22}, $\!\!$ 264.

\bibitem{Huang}
D.~{Huang}, P.~{Pintus}, C.~{Zhang}, Y.~{Shoji}, T.~{Mizumoto}, J.~E. {Bowers},
\newblock \emph{IEEE J. Sel. Top. Quant.} \textbf{2016}, \emph{22}, $\!\!$ 271.

\bibitem{Huang:17}
D.~Huang, P.~Pintus, Y.~Shoji, P.~Morton, T.~Mizumoto, J.~E. Bowers,
\newblock \emph{Opt. Lett.} \textbf{2017}, \emph{42}, $\!\!$ 4901.

\bibitem{HOM}
C.~K. Hong, Z.~Y. Ou, L.~Mandel,
\newblock \emph{Phys. Rev. Lett.} \textbf{1987}, \emph{59}, $\!\!$ 2044.

\bibitem{Yin:17}
Y.~Yin, Z.~Li, D.~Dai,
\newblock \emph{J. Lightwave Technol.} \textbf{2017}, \emph{35}, $\!\!$ 2227.

\bibitem{Hatori}
N.~Hatori, T.~Shimizu, M.~Okano, M.~Ishizaka, T.~Yamamoto, Y.~Urino, M.~Mori,
  T.~Nakamura, Y.~Arakawa,
\newblock \emph{J. Lightwave Technol.} \textbf{2014}, \emph{32}, $\!\!$ 1329.

\bibitem{8626148}
X.~Wang, X.~Quan, M.~Liu, X.~Cheng,
\newblock \emph{IEEE Photonics Technol. Lett.} \textbf{2019}, \emph{31}, $\!\!$
  349.

\bibitem{Xiang:20}
C.~Xiang, W.~Jin, J.~Guo, J.~D. Peters, M.~J. Kennedy, J.~Selvidge, P.~A.
  Morton, J.~E. Bowers,
\newblock \emph{Optica} \textbf{2020}, \emph{7}, $\!\!$ 20.

\bibitem{Harris}
N.~C. Harris, D.~Grassani, A.~Simbula, M.~Pant, M.~Galli, T.~Baehr-Jones,
  M.~Hochberg, D.~Englund, D.~Bajoni, C.~Galland,
\newblock \emph{Phys. Rev. X} \textbf{2014}, \emph{4}, $\!\!$ 041047.

\bibitem{LiuDajian}
D.~Liu, M.~Zhang, D.~Dai,
\newblock \emph{Opt. Lett.} \textbf{2019}, \emph{44}, $\!\!$ 1304.

\bibitem{Jin2}
H.~Jin, F.~M. Liu, P.~Xu, J.~L. Xia, M.~L. Zhong, Y.~Yuan, J.~W. Zhou, Y.~X.
  Gong, W.~Wang, S.~N. Zhu,
\newblock \emph{Phys. Rev. Lett.} \textbf{2014}, \emph{113}, $\!\!$ 103601.

\bibitem{Paesani}
S.~Paesani, Y.~Ding, R.~Santagati, L.~Chakhmakhchyan, C.~Vigliar, K.~Rottwitt,
  L.~K. Oxenløwe, J.~Wang, M.~G. Thompson, A.~Laing,
\newblock \emph{Nat. Physics} \textbf{2019}, \emph{15}, $\!\!$ 925.

\bibitem{Metcalf}
B.~J. Metcalf, J.~B. Spring, P.~C. Humphreys, N.~Thomas-Peter, M.~Barbieri,
  W.~S. Kolthammer, X.-M. Jin, N.~K. Langford, D.~Kundys, J.~C. Gates, B.~J.
  Smith, P.~G.~R. Smith, I.~A. Walmsley,
\newblock \emph{Nat. Photonics} \textbf{2014}, \emph{8}, $\!\!$ 770.

\bibitem{Knill}
E.~Knill, R.~Laflamme, G.~J. Milburn,
\newblock \emph{Nature} \textbf{2001}, \emph{409}, $\!\!$ 46.

\end{thebibliography}
\end{document}